%% file: klapp_5.tex
\documentclass[vecphys]{svmult}

\usepackage{makeidx}         
\usepackage{graphicx}        
\usepackage{multicol}        
\usepackage{cite}            
\usepackage{amssymb}
\usepackage{amsmath}  
\usepackage{colonequals}  
\usepackage{color}                 
\usepackage[bottom]{footmisc}

\makeindex             

\begin{document}

\title{Feedback control of colloidal transport}
\author{R.~Gernert \and S.~Loos \and K.~Lichtner \and S.~H.~L.~Klapp}
\institute{Institute of Theoretical Physics, 
Technische Universit\"at Berlin,
Hardenbergstra\ss e 36,
10623 Berlin, Germany
\texttt{klapp@physik.tu-berlin.de}}

\maketitle

\begin{abstract}
We review recent work on feedback control of one-dimensional colloidal systems, both with instantaneous feedback and with time delay. The feedback schemes are based
on measurement of the average particle position, a natural control target for an ensemble of colloidal particles, and the systems are investigated
via the \index{Fokker-Planck equation}Fokker-Planck equation for overdamped \index{Brownian particle}Brownian particles. Topics include the reversal of current and the emergence of current oscillations, 
transport in ratchet systems, and the enhancement of mobility by a co-moving trap. Beyond the commonly considered case of non-interacting systems,
we also discuss the treatment of colloidal interactions via \index{(dynamical) density functional theory}(dynamical) density functional theory and provide new results for systems with attractive interactions.
\end{abstract}

\section{Background}
\label{sec:1}
Within the last years, feedback control \cite{bechhoefer05} of colloidal systems,
that is, nano- to micron-sized particles in a thermally fluctuating bath of solvent particles, has become a focus of growing interest. Research in that area is stimulated, on the one hand,
by the fact that colloidal systems have established their role as theoretically and experimentally accessible model systems 
for equilibrium and nonequilibrium phenomena \cite{loewen13,Seifert,sagawa12} in statistical physics. Thus, colloidal systems are prime candidates to explore {\em concepts} of feedback control and its consequences. On the other hand, feedback control of colloidal particles has nowadays found its way into experimental applications.
Recent examples include control of colloids, bacteria and artificial motors in microfluidic set-ups \cite{qian13,bregulla14,braun14}, biomedical engineering \cite{fisher05}, and the manipulation of colloids by feedback traps \cite{cohen05,jun12,gieseler15}. Further, a series of recent experiments involving feedback control aims at exploring fundamental concepts of thermodynamics and information exchange in small stochastic systems \cite{toyabe10,jun14,gieseler15}.
As a consequence of these developments, feedback control of colloids is now an emerging field with relevance in diverse contexts, including optimization of self-asssembly processes \cite{juarez12}, and the manipulation of flow-induced behavior \cite{prohm14,strehober13} and rheology \cite{klapp10,vezirov15}.

Within this area of research, the present articles focuses on feedback control of {\em one-dimensional} (1D) \index{colloidal transport}colloidal transport. Transport in 1D systems without
feedback control has been extensively studied in the past decades, yielding a multitude of analytical and numerical results (see, e.g., \cite{haenggi09,Reimann,gernert14}). These have played a major role in understanding fundamentals of diffusion through complex landscapes and the role of noise. Paradigm examples of such 1D systems are Brownian particles driven through a periodic 1D ``washboard'' potential, or ratchet systems (\index{Brownian motor}Brownian motors) operating by a combination of asymmetric static potentials and time-periodic forces. It is therefore not surprising that the first applications of feedback control of colloids involve just these kinds of systems, pioneering studies being theoretical \cite{cao04,feito07,Craig08} and experimental \cite{lopez08}
investigations of a feedback-controlled 1D ``flashing ratchet''. Here it has been shown that the fluctuation-induced directed transport in the ratchet system can be strongly enhanced by
switching not under an externally defined, ``open-loop'' protocol, but with a closed-loop feedback scheme.

From the theoretical side, most studies focus on manipulating {\em single} colloidal particles (or an ensemble of non-interacting particles) in a 1D set-up, the basis being an overdamped or underdamped Langevin equation. The natural control target is then the position or velocity of the colloidal particle at hand.
Within this class, many earlier studies assume {\em instantaneous} feedback, i.e., no time lag between measurement and control action \cite{cao04}. However, there is now increasing interest in exploring systems with time delay \cite{feito07,Craig08,Craig,cao12,munakata14}. The latter typically 
arises from a time lag between the detection of a signal and the control action, an essentially omnipresent situation in experimental setups.
Traditionally, time delay was often considered as a perturbation; for example, in some ratchet systems it reduces the efficiency of transport \cite{feito07}. However, 
it is known from other areas that time delay can also have significant positive effects. For example, it can 
stabilize desired stationary states in sheared liquid crystals \cite{strehober13}, it can be used to probe coherent effects in electron transport in quantum-dot nanostructures
\cite{emary13}, and it can generate new effects such as current reversal \cite{hennig09,lichtner10} and spatiotemporal oscillations in extended systems \cite{lichtner12,gurevich13}. Moreover, 
time delay can have a {\em stabilizing} effect on chaotic orbits, a prime example being Pyragas' control scheme \cite{pyragas92}  of time-delayed feedback control \cite{schoellbuch}. 
Apart from the effects of time delay on the dynamical behavior, a further issue attracting increasing attention is the theoretical treatment of time-delayed, feedback-controlled (single-particle)
systems via stochastic thermodynamics \cite{munakata14,munakata09,Jiang11,munakata14,rosinberg15}.

Finally, yet another major question concerns the role of particle interactions. We note that, even in the idealized situation 
of a (dilute) suspension of non-interacting particles, feedback can induce {\em effective} interactions if the protocol involves system-averaged quantities \cite{cao04}.
For many real colloidal systems, however, direct interactions between the colloids stemming e.g., from excluded volume effects, charges on the particles' surfaces, or (solvent-induced) depletion effects cannot be neglected. Within the area of transport under feedback, investigations of the role of interactions have started only very recently.
Understanding the impact of interactions clearly becomes particularly important when one aims at feedback-controlling systems with phase transitions, pattern (or cluster-)forming systems, and systems with collective dynamic phenomena such as synchronization.

For an interacting, 1D colloidal system, one natural control variable is the {\em average} particle position, which is experimentally accessible e.g. by video microscopy. Theoretically,
the average position can be calculated from the time-dependent probability distribution $\rho(z,t)$, whose dynamics is determined by the Fokker-Planck (FP) equation \cite{risken_1984}
(for overdamped particles often called \index{Smoluchowski equation}Smoluchowski equation).
About two decades ago, Marconi and Tarazona \cite{marconi1,marconi2} have proposed a special type of FP equation, the so-called dynamical density functional theory, which is suitable for an interacting, overdamped system of colloidal particles. Within this framework, dynamical correlations are approximated adiabatically, and correlation effects enter via a free energy functional.

\begin{figure}[t]
\centering
\includegraphics*[width=.7\textwidth]{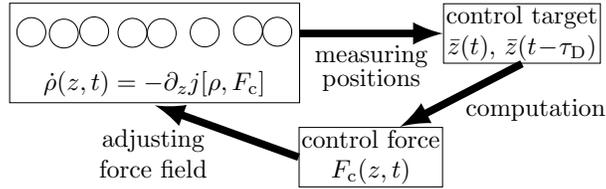}
\caption[]{Concept of feedback control for a system of (interacting) colloids. The control target is the average particle position $\bar{z}$ measured either at time $t$ or at a delayed time
$t-\tau_{\mathrm{D}}$. This average position determines the control force $F_{\mathrm{c}}(z,t)$. The system is investigated based on the Fokker-Planck equation where $\rho$ is the probability density and $j$ is the current.}
\label{control}       
\end{figure}

In this spirit, we have recently started to investigate a number of feedback-controlled 1D systems based on the FP formalism
\cite{lichtner10,lichtner12,loos_2014,gernert_2015}. The general scheme of feedback control used in these studies
is sketched in Fig.~\ref{control}.
The purpose of the present article is to summarize main results of these investigations.
We cover both, non-interacting systems and interacting systems, including new results for systems with attractive interactions.
Also, we discuss examples with instantaneous feedback and with time delay. We note that, in presence of time delay, the connection between the FP equation and the underlying Langevin equation is not straightforward (see, e.g., Refs.~\cite{Guillouzic99,Frank03,zeng12,munakata14,rosinberg15}), and this holds particularly for control schemes involving {\em individual} particle positions. 
However, here we consider the {\em mean} particle position as control target. For this situation, the results become consistent with those from a corresponding Langevin equation (with delayed force), if the number of realizations goes to infinity \cite{loos_2014}. 

\section{Theory}
\label{sec:theory}
We consider the motion of a system of $N$ overdamped colloidal particles at temperature $T$ in an external, one-dimensional, periodic potential $V_{\mathrm{ext}}(z)$ supplemented by a constant driving force $F_{\mathrm{ext}}$, where $z$ is the space coordinate. The particles are assumed to be spherical, with the size being characterized by the diameter $\sigma$.
In addition to thermal fluctuations, each particle experiences a time-dependent force $F_{\mathrm{c}}(z,t)$ which we will later relate to feedback control. We also allow for direct particle interactions which are represented by an interaction field $V_{\mathrm{int}}(z)$ to be specified later. 
The dynamics is investigated via the FP equation \cite{risken_1984} for the space- and time-dependent one-particle density
$\rho(z,t)=\langle \sum_{i=1}^N\delta(z-z_i(t))\rangle$ (where $\langle \dots\rangle$ denotes a noise average), yielding
\begin{eqnarray}
	\label{FPE}
	\partial_t \rho(z,t) & = &
			\partial_z\left[\gamma^{-1}\!\left(V_{\mathrm{ext}}'(z)\!-\!F_{\mathrm{ext}}\!-\!F_{\mathrm{c}}(z,t)\!+\!\partial_zV_{\mathrm{int}}(z,\rho)\right)\rho(z,t)\!+\!D_0\partial_z\rho(z,t)\right]
		\nonumber
	\\
	& = & -\partial_z j(z,t),
\end{eqnarray}
where $D_0$ is the short-time diffusion coefficient, satisfying the fluctuation-dissipation theorem \cite{risken_1984}
$D_0=k_{\mathrm{B}}T/\gamma$ (with $k_{\mathrm{B}}$ and $\gamma$ being the Boltzmann and the friction constant, respectively), and $j(z,t)$ is the probability current.
Throughout the paper, we measure the time $t$ in units of the ``Brownian'' time scale, $\tau_{\mathrm{B}}=\sigma^2/D_0$. For typical, $\mu$m-sized particles 
$\tau_{\mathrm{B}}$ is about $1$\,s \cite{lopez08,dalle11} or larger \cite{lee06}.

Feedback control is implemented through the time-dependent force $F_{\mathrm{c}}(z,t)$. Specifically, we assume this force to depend on the (time-dependent) average position
\begin{equation}
\label{z_average}
		\bar{z}(t)=\frac{1}{N}\int\mathrm{d}z\,\rho(z,t)\,z
		\,,
\end{equation}
where we have used that $N=\int \mathrm{d}z \,\rho(z,t)$. The density is calculated with periodic boundary conditions, that is,
$\rho(z+L_{\mathrm{sys}},t)=\rho(z,t)$ with $L_{\mathrm{sys}}$ being the system size. Thus, the time dependency of $F_{\mathrm{c}}(z,t)$ arises through the internal state of the system.

Our reasoning behind choosing the {\it mean} particle position rather 
than the individual position as 
control target is twofold: First, within the FPE treatment we have no access to the particle's position
for a given realization of noise, because the latter has already been averaged out. 
This is in contrast to previous studies using Langevin equations \cite{Craig,lopez08,Feito09} where the dynamical variable is the particle position itself. Second, the mean position is an experimentally accessible quantity, which can be monitored, e.g., by video microscopy \cite{Craig}.

\section{Non-interacting systems under feedback control}
\label{single}
\subsection{Particle in a co-moving \index{trap}trap}
\label{trap}
As a starting point \cite{gernert_2015}, we consider a single particle (or non-interacting colloids in a dilute suspension)
under the combined influence of a  static, \index{washboard potential}``washboard'' potential, 
\begin{equation}
V_{\mathrm{ext}}(z)=u(z)=u_0\sin^2(\pi z/a)
\label{V_ext}
\end{equation}
supplemented by a constant tilting force $F_{\mathrm{ext}}$ and the feedback force
\begin{equation}
F_{\mathrm{c}}(z,t)=-\partial_z V_{\mathrm{DF}}(z,t)
\end{equation}
derived from the potential
\begin{equation}
	V_\mathrm{DF}(z,t)=\eta (z-\bar{z}(t))^2
	\label{Vdf}
	\,.
\end{equation}
Physically speaking, Eq.~(\ref{Vdf}) describes a 
parabolic confinement, which moves instantaneously with the mean position, thus resembling the potential seen by particles in moving optical traps \cite{florin98,cole12}. The strength of the harmonic confinement, $\eta$, is set to constant.

In the absence of the potential barriers ($u_0=0$) the problem can be solved analytically. Starting from the initial condition $\rho(z,t\!=\!0)=\delta(z-z_0)$ one finds
$\bar{z}(t)=(F_{\mathrm{ext}}/\gamma)\,t+z_0$, yielding the mobility
\begin{equation}
\mu\colonequals \lim_{t\to\infty}\frac{\partial_t \bar{z}}{F_{\mathrm{ext}}}
=\frac{1}{\gamma}.
	\label{mobilityfree}
\end{equation}
Moreover, the mean-squared displacement describing the width of the distribution, 
\begin{equation}
w(t)=\langle (z-\bar{z}(t))^2\rangle
	\label{width}
\end{equation}
becomes
\begin{equation}
	w(t)=\frac{k_BT}{2\eta}\left(1-e^{-4\eta t/\gamma}\right),
	\label{width-free-diff}
\end{equation}
showing that density fluctuations {\em freeze} in the long-time limit. Interestingly, the same type of behavior of $w(t)$ occurs in a model of quantum
feedback control \cite{Brandes_2010}.

\begin{figure}[t]
\centering
\includegraphics*[width=.7\textwidth]{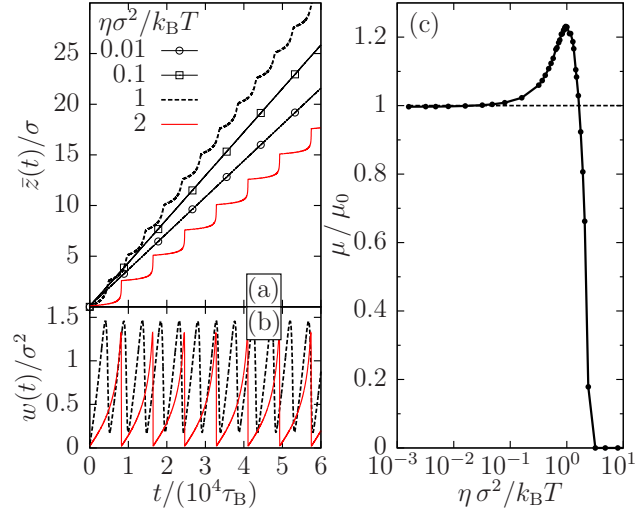}
\caption[]{Single particle in a trap \cite{gernert_2015}. (a) Average position and (b) width of the density distribution as functions of time. (c) Mobility (normalized by the mobility of the uncontrolled system)
as function of control strength.}
\label{trap_single}     
\end{figure}

For non-vanishing potential barriers and in presence of feedback,
 Eq.~(\ref{FPE}) has to be solved numerically. Figures~\ref{trap_single}(a) and (b) show representative results for the average position and the width.

Upon increase of $\eta$ the slope
of $\bar{z}(t)$ first increases but then decreases again. A further characteristic feature is the emergence of oscillations in $\bar{z}(t)$, the velocity $v(t)=\partial \bar{z}/\partial t$
and the width $w(t)$. These oscillations can be traced back to the periodic reconstruction of the effective energy landscape, 
$V_\mathrm{DF}(z,t)+u(z)$, which consists of a periodic increase and decrease of the energy barriers \cite{gernert_2015}. The period $\mathcal{T}$ of oscillations roughly coincides with the inverse \index{Kramers rate}Kramers rate
\cite{risken_1984,gernert14}, which is the relevant time scale for the slow barrier-crossing.
Also, the regime of pronounced oscillations partly coincides with the regime where a ``speed up'' of the motion occurs.
We quantify this ``speed up'' via an average mobility $\mu=\bar{v}/F_{\mathrm{ext}}$ based on the time-averaged velocity
$\bar{v}={\mathcal{T}}^{-1}\int_{t_1}^{t_1+\mathcal{T}}\mathrm{d}t\,v(t)$.
Figure \ref{trap_single}(c) shows $\mu/\mu_0$ depending on $\eta$, where $\mu_0\approx 1.2\cdot 10^{-4}/\gamma$ is the mobility of the uncontrolled system ($\eta\!=\!0$) with the same external potential \cite{stratonovich58,risken_1984}. For small $\eta$, we find $\mu\approx \mu_0$.
At intermediate values of $\eta$ the mobility shows a global maximum.
This maximum occurs in the range of $\eta$ where the oscillation periods of $v(t)$ are about (in fact, somewhat smaller than) the inverse Kramers rate.
For even larger values of $\eta$ one observes a sharp decrease of the mobility to zero. Here, the confinement induced by the trap becomes so strong that barrier diffusion is prohibited
(note that this effect would be absent if the trap was moved by an externally imposed velocity).
Overall, the increase of mobility by the co-moving trap is about twenty percent.
As we will see in Sec.~\ref{interactions}, a much more significant enhancement of mobility occurs when the particles interact.
\subsection{Feedback controlled \index{ratchet}ratchet}
\label{ratchet}
In the second example \cite{loos_2014},  $V_{\mathrm{ext}}(z)$ is a periodic, piecewise linear, ``sawtooth'' potential \cite{lopez08,Kamagawa98,Marquet02} 
defined by $V_{\mathrm{ext}}(z+a)=V_{\mathrm{ext}}(z)$ 
\begin{align}
	V_{\mathrm{ext}}(z)=
	\begin{cases}
		u_0z/(\alpha a), & 0<z \leq \alpha a,\\
		u_0z/((\alpha-1)a) , & (\alpha-1)a < z \leq 0
		\,,
	\end{cases}
	\label{Ratchet}
\end{align}
where $u_0$ and $a$ are again the potential height and the period, respectively, and $\alpha$ $\,\in[0,1]$ is the asymmetry parameter. The potential minimum within the central interval
$S=[(\alpha-1)a,\alpha a]$
is at $z=z_{\mathrm{min}}=0$. We further assume periodic boundary conditions such that $\rho(z+a,t)=\rho(z,t)$ (i.e., $a=L_{\mathrm{sys}}$), and we calculate
the mean position from Eq.~(\ref{z_average}) with the integral restricted to the interval $S$.

 In the absence of any further force ($F_{\mathrm{ext}}=0$ and $F_{\mathrm{c}}=0$) beyond that arising from $V_{\mathrm{ext}}(z)$, the system approaches for $t\to\infty$ an equilibrium state and thus there is no transport (i.e., no net particle current). It is well established, however, that by supplementing $V_{\mathrm{ext}}(z)$ by a time-dependent oscillatory force 
(yielding a ``rocking ratchet''), the system is permanently out of equilibrium 
and macroscopic transport can be achieved \cite{Reimann,Magnasco93,Bartussek}. 

Here we propose an alternative driving mechanism which is based on a {\it time-delayed} feedback force $F_{\mathrm{c}}(z,t)$ depending on the average particle position at an earlier time. Specifically, 
\begin{eqnarray} 
\label{f_FC}
F_{\mathrm{c}}(t) = -F\cdot \mathrm{sign} ( \bar{z}(t-\tau_{\mathrm{D}}) - z_0 )
\,,
\end{eqnarray}
where $\tau_\mathrm{D}$ is the delay time, $F$ is the amplitude (chosen to be positive), $z_0$ is a fixed position within the range $[0,\alpha a]$ (where $V_{\mathrm{ext}}$ 
increases with $z$), and the sign function
is defined by $\mathrm{sign}(x)=+1$ ($-1$) for $x>0$ ($x<0$). From Eq.~(\ref{f_FC}) one sees that the feedback force changes its sign whenever the delayed mean particle position $\bar{z}(t-\tau_{\mathrm{D}})$ becomes smaller or larger than $z_0$; we therefore call $z_0$ the ``switching'' position.

In the limit $\tau_\mathrm{D}\rightarrow 0$ any transport vanishes since the feedback force leads to a trapping of the particle at $z_0$.
This changes at $\tau_\mathrm{D}>0$. Consider a situation where the mean particle position at time $t$ is at the right side of $z_0$, while
it has been on the left side at time $t-\tau_{\mathrm{D}}$. 
In this situation the force $F_{\mathrm{c}}(t)$ points {\it away} from $z_0$ (i.e., $F_{\mathrm{c}}>0$), contrary to the case 
$\tau_{\mathrm{D}}=0$. Thus, the particle experiences a driving force towards the next potential valley, 
which changes only when the delayed position becomes larger than $z_0$. 
The force then points to the left until the delayed position crosses $z_0$ again. This oscillation of the force, together with the asymmetry of $V_{\mathrm{ext}}(z)$, creates a ratchet effect. 

\begin{figure}[t]
\centering
\includegraphics*[width=.7\textwidth]{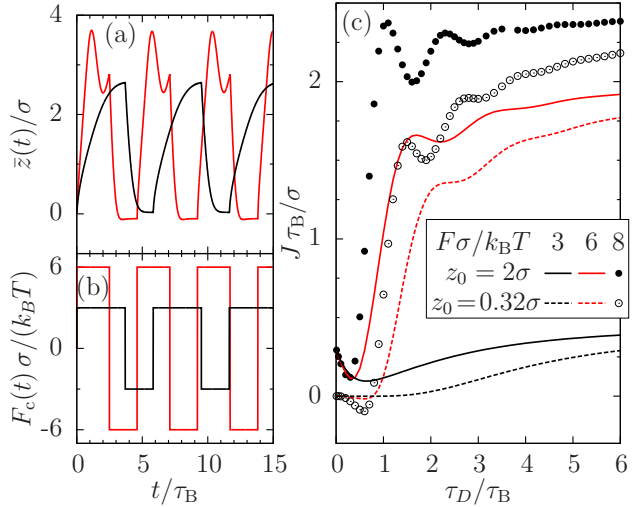}
\caption[]{Feedback-controlled ratchet \cite{loos_2014}.  (a) Average position and (b) feedback force
at two values of the force amplitude for delay time $\tau_{\mathrm{D}}=2\tau_{\mathrm{B}}$ and $z_0=2\sigma$.
(c) Average current (for different amplitudes and positions $z_0$ of the control force) as function of delay time.
Parameters are $L=8\sigma$ and $a=0.8$.}
\label{ratchet_position}      
\end{figure}

To illustrate the effect, we present in Fig.~\ref{ratchet_position}(a) exemplary data  for the time evolution of the mean particle position, $\bar{z}(t)$, which determines the control force. 
It is seen that
$\bar{z} (t)$ displays regular oscillations between values above and below $z_0$ for both force amplitudes considered. The period of these oscillations, $\mathcal{T}$, 
is roughly twice the delay
time. 
We note that the precise value of the period as well as the shape of the oscillations depend on the values of $F$ and $z_0$ \cite{loos_2014}.
Due to the oscillatory behavior of $\bar{z}(t)$ the delayed position $\bar{z}(t-\tau_{\mathrm{D}})$ oscillates around $z_0$ as well,
yielding a periodic switching of the feedback force between $+F$ and $-F$ with the same period
as that observed in $\bar{z} (t)$ [see Fig.~\ref{ratchet_position}(b)].
The oscillatory behavior of the feedback forces then induces a net current defined as
\begin{equation}
\label{J}
	J=\frac{1}{\mathcal{T}}\int_{t_1}^{t_1+\mathcal{T}}\mathrm{d}t'\,v(t')
\end{equation}
where $t_1$ is an arbitrary time after the ``equilibration'' period, $v(t)=\int_{S}\mathrm{d}z\,j(z,t)$ 
is the velocity, and $j(z,t)$ is calculated from the FPE~(\ref{FPE}) with periodic boundary conditions.
Numerical results for $J$ in dependence of the delay time $\tau_{\mathrm{D}}$ and the force amplitude
are plotted in Fig.~\ref{ratchet_position}(c).

The results clearly show that
the time delay involved in the feedback protocol is {\em essential} for the creation of a ratchet effect and, thus, for a nonzero net current.
For finite delay times ($\tau_{\mathrm{D}}\gtrsim 3\tau_{\mathrm{B}}$), the current generally increases with $\tau_{\mathrm{D}}$. Also, for a fixed $\tau_{\mathrm{D}}$, $J$ increases with increasing force amplitude (or with larger $z_0$).
At small delay times ($\tau_{\mathrm{D}}\lesssim 3\tau_{\mathrm{B}}$) the behavior of the function $J(\tau_{\mathrm{D}})$ is sensitive (in fact, behaves
non-monotonous) with respect to both, $F$ and $z_0$ \cite{loos_2014}.

Given the feedback-induced transport, it is interesting to 
compare the resulting current with
that generated by a conventional rocking ratchet. The latter is defined by replacing the force  $F_{\mathrm{c}}(t)$ in Eq. (\ref{FPE})
with a time-periodic (rectangular) force $F_{\mathrm{osc}}(t)=-F\cdot \mathrm{sign} \left[ \cos{ \left(\left( 2 \pi/T\right) t \right) } \right]$, where the period $T$ is set to the resulting period
$\mathcal{T}$ in the feedback-controlled case.
While the general behavior of the current (that is, small values of $J$ for small periods, saturation at large values for large periods) 
is similar for both, open-loop and closed-loop systems \cite{loos_2014}, the actual values of $J$ for a given period strongly depend on the type of control. 
It turns out that, for a certain range of switching positions (and not too
large delay times), the net current in the feedback-controlled system is actually {\em enhanced} relative to
the open-loop system.

A somewhat subtle aspect of the present model is that we introduce feedback on the level of the Fokker-Planck equation describing the evolution of the probability density. 
This is different from earlier studies 
based on the Langevin equation (see, e.g., \cite{Craig,lopez08,Feito09}), where the feedback is applied directly to
the position of one particle, $\chi_i(t)$, or to the average of $N$ particle positions $N^{-1}\sum_{i=1}^{N} \chi_i(t)$. Introducing feedback control in such systems
implies to introduce effective {\em interactions} between the particles. As a consequence, 
the transport properties in these particle-based models depend explicitly on the number of particles, $N$. From the perspective of these Langevin-based models,
the present model corresponds to the ``mean-field'' limit $N\to\infty$ (for a more detailed discussion, see \cite{loos_2014}).

\section{Impact of particle interactions}
\label{interactions}
We now turn to (one-dimensional) transport in systems of interacting colloids. To construct the corresponding contribution   
$V_{\mathrm{int}}(z)$ in the FP equation (\ref{FPE}), we employ concepts
from dynamical density functional theory (DDFT)
\cite{marconi1,marconi2,archer04}. Within the DDFT, the exact FP equation for an overdamped system with (two-particle) interactions
is approximated such that non-equilibrium two-particle correlations
at time $t$ are set to those of an equilibrium system with
density $\rho(z,t)$. This {\em adiabatic approximation} allows to formally relate the interaction contribution to the FPE to the excess free energy of an equilibrium system
[whose density profile $\rho_{\mathrm{eq}}(z)$ coincides with the instantaneous density profile $\rho(z,t)$]. It follows that
\begin{equation}
V_{\mathrm{int}}(z)=\frac{\delta{\cal F}^{\mathrm{int}}[\rho]}{\delta\rho(z,t)}
\label{vint}
\end{equation}
where ${\cal F}^{\mathrm{int}}[\rho]$ is the excess (interaction) part of the {\em equilibrium} free energy functional. Thus, one can use well-established 
equilibrium approaches as an input
into the (approximate) dynamical equations of motion.
\subsection{\index{current reversal}Current reversal}
Our first example involves ``ultra-soft'' particles interacting via the Gaussian core potential (GCM) 
\begin{equation}
	v_\mathrm{GCM}(z,z')=\varepsilon\exp\left(-\frac{(z-z')^2}{\sigma^2}\right),
	\label{vgcm}
\end{equation}
(with $\varepsilon>0$), a typical coarse-grained potential modeling a wide class of soft, partially penetrable macroparticles (e.g., polymer coils)
with effective (gyration) radius $\sigma$ \cite{likos01,louis00}. Due to the penetrable nature of the Gaussian potential which allows an, in principle, infinite number of neighbors,
the equilibrium structure of the GCM model can be reasonably calculated within the mean field (MF) approximation
\begin{equation}
{\cal F}^{\mathrm{int}}[\rho]=\frac{1}{2}\int d z\int d z'\rho(z,t)v^{\mathrm{GCM}}(z,z')\rho(z',t).
\label{meanfield}
\end{equation}
The MF approximation
is known to become quasi-exact in the high-density limit and yields reliable results
even at low and moderate densities \cite{louis00}. 

The particles are subject to an external {washboard potential} of the form defined in Eq.~(\ref{V_ext}) plus a constant external force $F_{\mathrm{ext}}=3 k_{\mathrm{B}}T / \sigma$.
To implement feedback control we use the time-delayed force 
\begin{equation}
\label{f_control1}
F_{\mathrm{c}}(z,t)= F_{\mathrm{c}}(t)=-K_0\left (1-\tanh\left[\frac{N}{\sigma}\left(\bar{z}(t)-\bar{z}(t-\tau_{\mathrm{D}})\right)\right]\right),
\end{equation}
which involves the difference between the average position at times $t$ and $t-\tau_{\mathrm{D}}$. 
By construction, $F_{\mathrm{c}}(t)$ vanishes in the absence of time delay ($\tau_{\mathrm{D}}=0$).
The idea to use a feedback force depending on the difference of the control target at two times 
is inspired by the time-delayed feedback control method suggested by Pyragas \cite{pyragas92} in the context of chaos control. Indeed, the original idea put foward by Pyragas was to stabilize certain unstable {\em periodic} states in a non-invasive way (notice that $F_{\mathrm{c}}(t)$ vanishes if $\bar{z}(t)$ performs periodic motion with period $\tau_{\mathrm{D}}$). Later,
Pyragas control has also been used to stabilize steady states (for a recent application in driven soft systems, see \cite{strehober13}). We also note that a similar strategy has been used on the level of an (underdamped) Langevin equation by Hennig {\em et al.} \cite{hennig1}.

\begin{figure}
\centering
\includegraphics*[width=.7\textwidth]{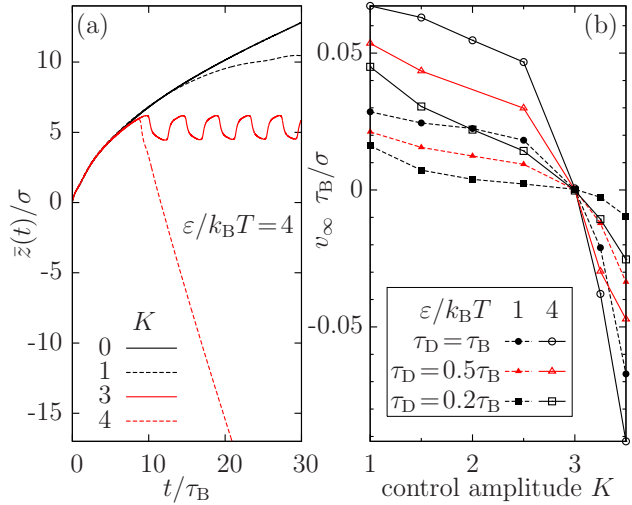}
\caption{Feedback-controlled particle in a washboard potential with tilt \cite{lichtner10}.
(a) Average particle position as a function of time for various control amplitudes $K$. (b) Long-time velocity as function
of control amplitude for various coupling strengths and time delays.}
\label{current1}
\end{figure}

The impact of the control force $F_{\mathrm{c}}(t)$ on the average particle position $\bar{z}(t)$ is illustrated in Fig~\ref{current1},
where we have chosen a moderate value of the driving force (yielding rightward motion in the uncontrolled system) and a delay time
equal to the ``Brownian'' time, $\tau_{\mathrm{D}}=\tau_{\mathrm{B}}$. 
In the absence of control ($K=K_0\sigma/k_{\mathrm{B}}T=0$) the average position 
just increases with $t$ reflecting rightward motion, as expected.
The slope of the function $\bar{z}(t)$ at large $t$ may be interpreted as the long-time velocity
$v_{\infty}=\lim_{t\to\infty} d\bar{z}(t)/dt$.
Increasing $K$ from zero, the velocity first decreases until the motion stops (i.e., the time-average of
$\bar{z}(t)$ becomes constant) at $K=3$. This value corresponds to a balance between control force and biasing driving force.
Here, the average position $\bar{z}(t)$ displays an {\em oscillating} behavior changing
between small backward motion and forward motion, with a period of about $5\tau_{\mathrm{B}}$ (that is, much larger than the delay time).
These oscillations are accompanied by oscillations of the {\it effective} force
$F^{\mathrm{eff}}=F_\mathrm{c}(t)+F_\mathrm{ext}$ around zero
(notice the restriction
$-2K_0\leq F_{\mathrm{c}}(t)\leq 0$). Consistent with this observation, there is no directed net motion.
A more detailed discussion of the onset of oscillations is given in Ref.~\cite{lichtner12}, where we have focussed on a non-interacting system
($\epsilon=0$). Indeed, for the present situation we have found that a non-interacting ensemble 
subject to the Pyragas control (\ref{f_control1}) behaves qualitatively similar to its interacting counterpart.
Moreover, for the non-interacting case, we have identified
the onset of oscillations as supercritical Hopf bifurcation.

Turning back to Fig.~\ref{current1}(a) we see that even larger control amplitudes ($K>3$) result in a significant backward motion, 
i.e., $\bar{z}(t)$ and $v_{\infty}$ become negative. Thus, the feedback control induces
{\em current reversal}.

To complete the picture, we plot in Fig.~\ref{current1}(b) the long-time velocity $v_{\infty}$ (averaged over the oscillations of $\bar{z}(t)$, if present)
as function of the
control amplitude. We have included data for different delay times $\tau_{\mathrm{D}}$ and different interaction (i.e., repulsion) strengths $\varepsilon$.
All systems considered display a clear current reversal at $K=3$ (balance between feedback and bias), 
where the velocity $v_{\infty}$ changes from positive to negative values irrespective of $\varepsilon$ and $\tau_{\mathrm{D}}$.
Regarding the role of the delay we find that, at fixed coupling strength $\varepsilon$, $v_{\infty}$ decreases in magnitude when the delay time decreases
from $\tau_{\mathrm{D}}=\tau_{\mathrm{B}}$ towards $\tau_{\mathrm{D}}=0.2\tau_{\mathrm{B}}$. 
In other words, the time delay {\em supports} the current reversal in the parameter range considered.
Regarding the interactions, Fig.~\ref{current1}(b) shows that reduction of $\varepsilon$ (at fixed $\tau_{\mathrm{D}}$) yields a decrease of the magnitude of
$v_{\infty}$ as compared
to the case $\varepsilon/k_{\mathrm{B}}T=4$. Thus, repulsive interactions between the particles yield a ``speed up'' of motion.
\subsection{Interacting particles in a trap}
As a second example illustrating the impact of particle interactions we turn back to the feedback setup discussed in Sec.~\ref{trap},
that is, feedback via a co-moving harmonic trap. In Sec.~\ref{trap} we have discussed this situation for a single colloidal particle driven through a washboard potential.
In that case, feedback leads to a slight, yet no dramatic increase of the transport efficiency as measured by the mobility. 

\begin{figure}
	\centering
	\includegraphics[width=.7\linewidth]{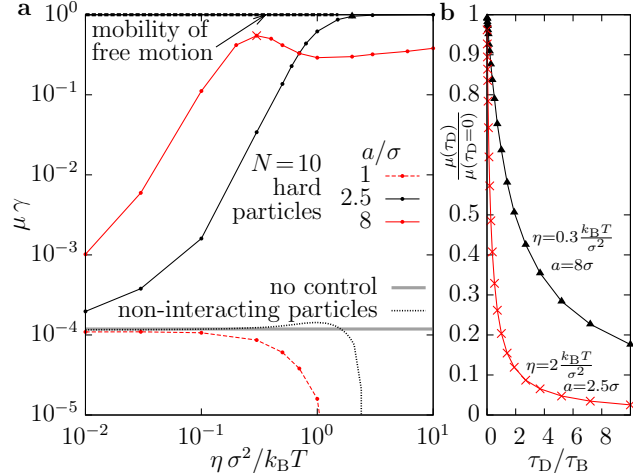}
	\caption{(a) Mobility of a system of hard particles in dependence of the control strength, $\eta$, for various values of the lattice constant. 
	The mobility can rise up to $1/\gamma$, the mobility of free motion.
	The thick line indicates the mobility in the uncontrolled case. (b) Impact of time delay at the parameters indicated by a triangle and cross in part (a).}
	\label{mobility_hard}
\end{figure}

This changes dramatically when the particles interact. In \cite{gernert_2015} we have explored the effect of two types of repulsive particle interactions, one of them
being the Gaussian core potential introduced in Eq.~(\ref{vgcm}). Here we focus on results for hard particles described by the interaction potential
\begin{gather}
	v_\mathrm{hard}(z,z')=\begin{cases} 0 &,\;\text{for}\;\vert z-z'\vert \ge \sigma \\\infty&,\;\text{for}\;\vert z-z'\vert<\sigma \end{cases}
	\label{vhc}
	\,.
\end{gather}
For one-dimensional systems of \index{hard spheres}hard spheres there exists an exact free energy functional \cite{percus76} derived by Percus, which corresponds to the 
one-dimensional limit of fundamental measure theory \cite{roth10}. This functional is given by
\begin{equation}
	{\cal F}^\mathrm{int}[\rho]=-\frac{1}{2}\int\mathrm{d}z\,\ln\left(1-\ell[\rho,z,t]\right)
[\rho(z+\frac{\sigma}{2},t)+\rho(z-\frac{\sigma}{2},t)]
	\label{Fhc}
	\,,
\end{equation}
where
\begin{equation}
	\ell[\rho,z,t]=\int_{z-\sigma/2}^{z+\sigma/2}\mathrm{d}z'\,\rho(z',t)
	\label{localpackingfraction}
\end{equation}
is the local packing fraction. Corresponding results for the mobility are shown in Fig.~\ref{mobility_hard}(a).
For appropriately chosen lattice constants ($a>\sigma$), we observe a dramatic increase of $\mu$ with $\eta$ and $N$ over several orders of magnitude. This is in striking contrast 
to the corresponding single-particle result (see dotted line in Fig.~\ref{mobility_hard}(a)), and similar behavior occurs
for ultra-soft particles \cite{gernert_2015}. In fact, for specific values $\eta$ and $N$, the mobility increases up to the maximal possible value $\mu=1/\gamma$, 
the mobility of free (overdamped) motion. 

The dramatic enhancement of transport can be understood by considering the (time-dependent) energy landscape formed 
by the combination of external potential $u(z)$, feedback potential $V_\mathrm{DF}(z,t)$ [see Eq.~(\ref{Vdf})]
and interaction contribution $V_\mathrm{int}(z)$ \cite{gernert_2015}. It turns out that the
$V_{\mathrm{int}}(z,t)$ develops peaks at the minima of the potential $V_\mathrm{DF}+u$. The interaction contribution thus
tends to ``fill'' the valleys, implying that the energy barriers between the minima decrease. This
results in an enhancement of diffusion over the barriers and thus, to faster transport. In other words, interacting particles ``help each other'' to overcome the external barriers.

\paragraph{Delayed trap}
Given that any experimental setup of our feedback control involves a finite time to measure the control target (i.e., the mean position), we
briefly consider the impact of time delay. To this end we change the control potential defined in Eq.~\eqref{Vdf} into the expression
\begin{equation}
	V_\mathrm{DF}^\mathrm{delay}(z,\rho)=\eta\,(z-\bar{z}(t-\tau_\mathrm{D}))^2
	\label{Vdfdelay}
	\,.
\end{equation}
We now consider two special cases involving hard particles, where the non-delayed feedback control leads to a particularly high mobility. Numerical results are plotted in 
Fig.~\ref{mobility_hard}(b),
showing that the delay causes a pronounced decrease of mobility. To estimate
the consequences for a realistic colloidal system, we note that
feedback mechanisms can be implemented at the time scale of $10$\,ms \cite{lopez08,cohen06,bregulla14} where $\tau_B$ (the timescale of Brownian motion) is for 
$\mu$m sized particles of the order of $1$\,s \cite{lopez08,dalle11} or larger \cite{lee06}. Hence, we expect that the ratio $\tau_\mathrm{D}/\tau_B$ is rather small, that is, of the order $10^{-1}$. For such situations, our results in Fig.~\ref{mobility_hard}(b) predict only a small decrease of $\mu$ relative to the non-delayed case. 

\paragraph{Attractive interactions}
Given the strong enhancement of mobility it clearly is an interesting question to which extent these observations depend on the type of the interactions. In \cite{gernert_2015} we have observed very similar behavior for two, quite different types of {\em repulsive} interactions. What would happen in presence of additional attractive interactions? 

Indeed, in colloidal systems attractive forces quite naturally arise through the so-called depletion effect, which originates from the large size ratio between the colloidal and the
solvent particles: when two colloids get so close that solvent particles do not fit into the remaining space, the accessible volume of the colloids effectively increases, yielding
a short-range ``entropic'' attraction with a range determined by the solvent particles' diameter.
Other sources of attraction are van-der-Waals forces \cite{bishop09}, or the screened Coloumbic forces between oppositely charged colloids \cite{leunissen05}. 
A generic model to investigate the impact of attractive forces between colloids is the hard-core attractive Yukawa (HCAY) model \cite{caccamo96} defined by
\begin{equation}
	v_\mathrm{hcay}(z,z')=v_\mathrm{hard}(z,z')-Y\frac{\exp(\kappa(\sigma-\vert z-z'\vert))}{\vert z-z'\vert/\sigma},
	\label{vhcay}
\end{equation}
where $v_\mathrm{hard}(z,z')$ has been defined in Eq.~(\ref{vhc}), and
the parameters $Y$ and $\kappa$ determine the strength and range of the attractive part, respectively.
Here we set $Y/k_\mathrm{B}T=10$ and consider the range parameters $\kappa\sigma=7$ and $\kappa\sigma=1$. 
The former case refers to a typical depletion interactions (whose range is typically much smaller than the particle diameter) \cite{hagen94,dijkstra02,mederos94},
whereas the second case rather relates to screened Coloumb interactions. In both cases, the {\em three-dimensional} HCAY system
at $Y/k_\mathrm{B}T=10$ would be phase-separated (gas-solid coexistence) \cite{dijkstra02}. In other words, our choice of $Y$ corresponds to a strongly
correlated situation. To treat the HCAY interaction within our theory, we construct a corresponding potential [see Eq.~(\ref{vint})]
from the derivative of the (exact) hard-sphere functional given in Eq.~(\ref{Fhc}) 
combined with the mean-field functional (\ref{meanfield}) for the Yukawa attraction.

\begin{figure}
	\centering
	\includegraphics[width=.7\linewidth]{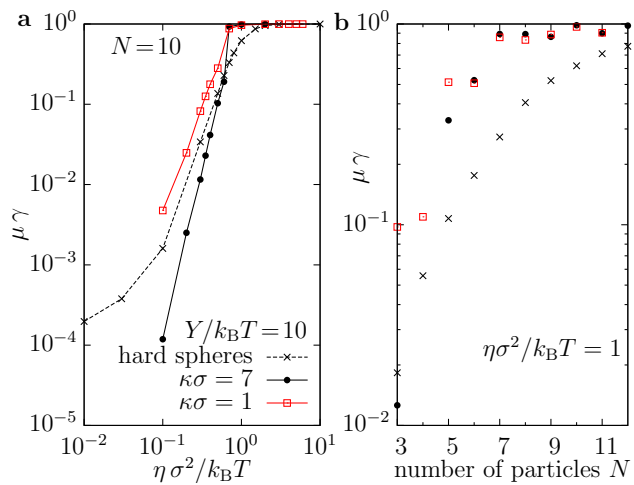}
	\caption{(a) Mobility of a system of hard particles with additional short-ranged attractive interactions 
	 as function of the control strength, $\eta$, and two range constants. (b) Impact of particle number.}
	\label{mobility_soft}
\end{figure}

Numerical results for the mobility of the (one-dimensional) HCAY system under feedback control are 
plotted in Fig.~\ref{mobility_soft}(a) together with corresponding results for the (purely repulsive) hard sphere system. 
The general dependence 
of the mobility on the feedback strength seems to be quite insensitive to the detail of interactions: In all three cases we find an enhancement of
$\mu$ towards the value characterizing a freely (without barriers) diffusing particle. Quantitatively, the results in the range
$\eta\sigma^2 \lesssim 0.7k_\mathrm{B}T$ depend on the range parameter $\kappa$. In particular, the system with the longer range of attraction
($\kappa\sigma=1$)
has a higher mobility than the one at $\kappa\sigma=7$, with the mobility of the second one being even smaller than that in the hard-sphere system.
However, at $\eta\sigma^2 \ge 0.7k_\mathrm{B}T$ both HCAY mobilities exceed the hard-sphere mobility. The physical picture is that of a moving ``train'' of particles, where
each particle not only pushes its neighbors (such as in the repulsive case) but also drags them during motion.

Finally, we consider in Fig.~\ref{mobility_soft}(b) the dependence of the mobility on the total number of particles, $N$ (at fixed feedback strength $\eta$). This dependence arises
from the fact that the length of the particle ``train'', $N\sigma$, competes with the two other relevant length scales, that is, the effective
size of the trap (controlled by $\eta$), and the wavelength $a$. Thus, increasing $N$ in the presence of particle interactions means to ``compress'' the train. For all systems
considered in Fig.~\ref{mobility_soft}(b) this compression leads to an increase of mobility since, as shown explicitely in \cite{gernert_2015} for hard-sphere systems, the barriers
in the {\em effective} potential landscape become successively smaller. From Fig.~\ref{mobility_soft}(b) we see that 
the increase of $\mu$ with $N$ is even more pronounced in presence of colloidal attraction, suggesting that attractive forces enhance the rigidity of the train.

\section{Conclusions and outlook}
In this article we have summarized recent research on feedback control in 1D colloidal transport. 
We close with pointing out some open questions
and possible directions for future research. 

A first notion concerns the role of the control target and the theoretical formalism employed. The (Fokker-Planck based) approach described in Secs.~\ref{sec:theory} - \ref{interactions}
assumes control schemes targeting the {\em average} particle position, which seems to be the natural, i.e., experimentally accessible, choice for a realistic system of (interacting) colloids. Moreover, the FP approach allows for a convenient treatment of colloidal interactions via the DDFT approach, which have been typically neglected in earlier, (Langevin-based) investigations.
However, it remains to be clarified how the FP results relate to findings from 
Langevin-based  investigations targeting the {\em individual} positions (or other degrees of freedom), which is the straightfoward
way to control a {\em single} colloidal particle. In other words, in which respect does an ensemble of colloids behave differently from a single one under feedback control?
These issues become particularly dramatic in the case of time-delayed feedback control, where the Langevin
equation is non-Markovian and the FP description consists, in principle, of an infinite hierarchy of integro-differential equations (see, e.g., \cite{rosinberg15})
We note that even if one takes the average position as control target on the Langevin level, the results become consistent with those from our FP approach only in the
limit $N\to\infty$ \cite{loos_2014}.

Conceptual questions of this type are also of importance in the context of stochastic thermodynamics. As pointed out already in Sec.~\ref{sec:theory}
there is currently a strong interest (both in the classical and in the quantum systems community)
to explore the role of feedback for the exchange of heat, work and entropy of a system with its environment \cite{munakata14,munakata09,Jiang11,rosinberg15}. 
This is usually done by considering 
the entropy production, second-law like inequalities and fluctuation relations. In \cite{loos_2014}, we have presented some numerical results for the
entropy production in the time-delayed feedback controlled rocking ratchet described in Sec.~\ref{single}, the goal being to evaluate the efficiency
of feedback control versus open-loop control. However, 
systematic investigations of feedback systems with time delay are just in their beginnings. This is even more true for systems with direct (pair) interactions. 

A further interesting question from a physical point of view concerns the role of spatial dimension. In the present article we have focused (as it is mostly done) on 1D systems.
Clearly, it would be very interesting to develop feedback control concepts for two-dimensional, interacting colloidal systems where, in addition to particle chain and cluster formation,
anisotropic collective transport mechanisms \cite{pedrero2015}, phase transitions \cite{archer2013}, spinodal decomposition, and more complex pattern formation such as stripe formation 
\cite{lichtner2014} can occur. 
From the perspective of the present theoretical approach, which is based on the FP equation, 
a main challenge for the 2D case arises through the fact that we handle interaction effects on the basis of dynamical density functional theory (DDFT). For example, contrary to the 1D case there is
no exact functional for hard spheres in two dimensions, making the entire approach less accurate. Thus, it will become even more important to test any FP-DDFT results against
particle-resolved (Brownian Dynamics) simulations.  One distinct advantage of the FP-DDFT approach, however, is that one can perform further approximations 
such as gradient expansions. This would 
allow to establish a relation to the large amount of work on feedback-controlled pattern forming systems based on (continuum) partial differential equations (see, e.g., \cite{schoellbuch,gurevich13}).

Finally, we want to comment on the experimental feasibility of our feedback protocols. To this end we first note that
state-of-the-art video microscopy techniques allow to monitor particles as small as 20 nm \cite{selmke14}. This justifies the use of (average) particle positions as control targets for 
colloids with a broad range of sizes from the nanometer to the micron scale.
Typical experimental delay times (arising from the finite time required for particle localisation) are about 5-10 ms for single particles (see, e.g., \cite{bregulla14,jun14}). These values are substantially smaller than typical diffusion (``Brownian'') time scales ($\approx 500$ms - $1\mu$s), which underlines the idea that the relative time delay
in colloidal transport is typically small. Naturally, somewhat larger delay times
are expected to arise in feedback control of several (interacting) particles. Still, we think that our feedback protocols for many-particle systems are feasible, last but not least
because many-particle monitoring techniques are being continuously improved \cite{gomez-solano15}.
We thus hope that the recent theoretical advancements reported in this article and in related theoretical studies will stimulate further experimental work.

\section*{Acknowledgements}
This work was supported by the Deutsche Forschungsgemeinschaft through SFB 910 (project B2).

%
\input{referenc}



\printindex
\end{document}

%% file: referenc.tex
%
%

%
%